\documentstyle[twoside]{article}
\begin{document}

\title{TROUBLES WITH QUANTUM ANISOTROPIC COSMOLOGICAL MODELS:
LOSS OF UNITARITY}

\author{F.G. Alvarenga\thanks{%
e-mail: flavio@cce.ufes.br}, A.B. Batista\thanks{%
e-mail: brasil@cce.ufes.br},
J. C. Fabris\thanks{%
e-mail: fabris@cce.ufes.br}, S.V.B. Gon\c{c}alves\thanks{%
e-mail: sergio@cce.ufes.br}\\
Departamento de F\'{\i}sica, Universidade Federal do Esp\'{\i}rito Santo, Brazil}
\maketitle
\abstract{
The anisotropic Bianchi I cosmological model coupled with perfect fluid is
quantized in the minisuperspace. The perfect fluid is described by using the
Schutz formalism which allows to attribute dynamical degrees of freedom to
matter. It is shown that the resulting model is non-unitary. This breaks
the equivalence between the many-worlds and dBB interpretations of quantum
mechanics.}

\vspace{0.5cm}

PACS number(s): 04.20.Cv., 04.20.Me
\vspace{0.5cm}

The aim of this work is to study the anisotropic Bianchi I quantum cosmological
model coupled to a perfect fluid described by the Schutz formalism\cite{schutz1,schutz2}.
Hence, we start with the classical Lagrangian given by
\begin{eqnarray}
\label{action}
{\cal A} &=& \int_Md^4x\sqrt{-g}R + 2\int_{\partial M}d^3x\sqrt{h}h_{ab}K^{ab}
\nonumber\\
&+& \int_Md^4x\sqrt{-g}p 
\end{eqnarray}
where $K^{ab}$ is the extrinsic curvature, and $h_{ab}$ is the induced
metric over the three-dimensional spatial hypersurface, which is
the boundary $\partial M$ of the four dimensional manifold $M$; the
factor $16\pi G$ is made equal to one.
The first two terms were first obtained in the classical work of
Arnowitt, Deser and Misner\cite{adm};
the last term of (\ref{action}) represents the matter contribution to
the total action in the Schutz's formalism for perfect fluids, $p$ being the pressure, which is linked to the energy density by the equation of state $p = \alpha\rho$. In the
Schutz's formalism\cite{schutz1,schutz2},
the four-velocity is expressed in terms of five potentials $\epsilon$,
$\zeta$, $\beta$, $\theta$ and $S$:
\begin{equation}
U_\nu = \frac{1}{\mu}(\epsilon_{,\nu} + \zeta\beta_{,\nu} +
\theta S_{,\nu})
\end{equation}
where $\mu$ is the specific enthalpy. The variable $S$ is the specific
entropy, while the potentials $\zeta$ and $\beta$ are connected with
rotation and are absent for FRW's type models. The variables $\epsilon$ and
$\theta$ have no clear physical meaning.
The four velocity is subject to the condition
\begin{equation}
U^\nu U_\nu = 1 \quad .
\end{equation}
\par
The metric describing a Bianchi I anisotropic model is given by
\begin{equation}
ds^2 = N^2dt^2 - \biggr(X^2dx^2 + Y^2dy^2 + Z^2dz^2\biggl)\quad .
\end{equation}
In this expression, $N$ is the lapse function.
Using the constraints for the fluid, and after some thermodynamical considerations,
the final reduced action, where surface terms were discarded,
takes the form
\begin{eqnarray}
&{\cal A}& = \int dt\biggr[-\frac{2}{N}\biggr(\dot X\dot YZ + \dot X\dot ZY +
\dot Y\dot ZX\biggl)
+ N^{-1/\alpha}\nonumber\times\\
&\times&(XYZ)\frac{\alpha}{(\alpha + 1)^{1/\alpha + 1}}(\dot\epsilon +
\theta\dot S)^{1/\alpha + 1}e^{(- \frac{S}{\alpha})}
\biggl] \quad ,
\end{eqnarray}
where $\alpha$ defines the equation of state of the matter field: $p = \alpha\rho$.
\par
At this point, is more suitable to redefine the metric coefficients as
\begin{eqnarray}
&X&(t) = e^{\beta_0 + \beta_+ + \sqrt{3}\beta_-}\quad ,\quad Y(t) = e^{\beta_0 + \beta_+ - \sqrt{3}\beta_-}\quad ,\nonumber\\
&Z&(t) = e^{\beta_0 - 2\beta_+} \quad . 
\end{eqnarray}
Using these new variables, the action may be simplified further, leading
to the gravitational Lagrangian density
\begin{equation}
L_G = -6\frac{e^{3\beta_0}}{N}\{\dot\beta_0^2 - \dot\beta_+^2 - \dot\beta_-^2\} \quad .
\end{equation}
\par
From this expression, we can evaluate the conjugate momenta:
\begin{eqnarray}
&p&_0 = - 12\frac{e^{3\beta_0}}{N}\dot\beta_0 \quad , \quad p_+ = 12\frac{e^{3\beta_0}}{N}\dot\beta_+ \quad ,\nonumber\\
&p&_- = 12\frac{e^{3\beta_0}}{N}\dot\beta_- \quad .
\end{eqnarray}
\par
The matter sector may be recast in a more suitable form through the canonical transformations
\begin{eqnarray}
T = p_Se^{-S}p_\epsilon^{-(\alpha + 1)} \quad , \quad 
p_T = p_\epsilon^{\alpha + 1}e^S \quad ,\nonumber\\
\bar\epsilon = \epsilon - (\alpha + 1)\frac{p_S}{p_\epsilon} \quad ,
\quad \bar p_\epsilon = p_\epsilon \quad .
\end{eqnarray}
The final expression for the total Hamiltonian is
\begin{equation}
\label{hamilton}
H = N\biggr\{\frac{e^{-3\beta_0}(- p_0^2 + p_+^2 + p_-^2)}{24} + e^{-3\alpha\beta_0}p_T\biggl\} .
\end{equation}
\par
Standard quantization in the mini-superspace, using canonical procedures, leads to the Schr\"odinger-like equation
\begin{eqnarray}
\label{wdwe}
\biggr(\frac{\partial^2}{\partial\beta^2_0} - \frac{\partial^2}{\partial\beta^2_+} -
\frac{\partial^2}{\partial\beta^2_-}\biggl)\psi = 
- 24ie^{3(1 - \alpha)\beta_0}\frac{\partial\psi}{\partial T}.
\end{eqnarray}
Note that the variable associated with the matter field may be identified with time,
since it appears linearly in the Hamiltonian. Another important feature of this
Schr\"odinger-like equation, as it will be discussed later, is the hyperbolic signature
of the its kinetic part.
\par
This equation can be solved by using the method of separation of variables, leading to the
following final expression for the wave function:
\begin{eqnarray}
\label{sol}
\Psi &=& e^{i(k_+\beta_+ + k_-\beta_-)}\biggr[\bar C_1J_\nu\biggr(\frac{\sqrt{24E}}{r}a^r\biggl) 
\nonumber\\&+&
\bar C_2J_{-\nu}\biggr(\frac{\sqrt{24E}}{r}a^r\biggl)\biggl]e^{-iET}
\end{eqnarray}
where $k_\pm$ are separation constants, $E$ defines the energy of the
stationary states and $\bar C_{1,2}$ are integration constants. We have also
defined $a = e^{\beta_0}$ and $r = \frac{3}{2}(1 - \alpha)$.
\par
We construct a wave packet for the simplified case where $k_0$.
The wave packet is then given by
\begin{equation}
\Psi = \int \int A(k_+,q)e^{ik_+\beta_+}J_\nu(qa^r)e^{-iq^2T}dk_+dq  ,
\end{equation}
with $q = \frac{\sqrt{24E}}{r}$ and
\begin{equation}
\label{sup}
A(k_+,q) = e^{-\gamma k_+^2}q^{\nu + 1}e^{-\lambda q^2} \quad .
\end{equation}
In this case, the integrals can be explicitly calculated, leading to the expression
\begin{equation}
\label{wp}
\Psi = \frac{1}{B}\sqrt{\frac{\pi}{\gamma}} \exp\biggr[- \frac{a^{2r}}{4B} - \frac{(\beta_+  + C(a,B))^2}{4\gamma}\biggl]
\end{equation}
where
\begin{eqnarray}
B &=& \lambda + isT \quad , \quad C(a,B) = \ln a - \frac{2}{3(1 - \alpha)}\ln 2B \quad ,\nonumber
\\
s &=& - \frac{3(1 - \alpha)^2}{32} \quad .
\end{eqnarray}
Notice that the wave packet given by (\ref{wp}) is square integrable, and it vanishes in the extremes of the interval of validity of
the variables $a = e^{\beta_0}$ and $\beta_+$, except along the line $\beta_0 = - \beta_+$ where it
takes a constant value, being consequently regular as it is physically required.
The wave packet (\ref{wp}) is indeed a solution of the equation (\ref{wdwe}), as it can
be explicitly verified, and it obeys the boundary conditions fixed before. If we discard
the terms corresponding to the variable $\beta_+$(connected with $k_+$), the wave packet for
the isotropic case\cite{nivaldo1} is reobtained.
\par
The main point to be remarked now is that the norm of (\ref{wp}) is time dependent.
Using the definition $a = e^{\beta_0}$, integrating in
$\beta_+$ and $a$, taking into account an unusual measure in order to preserve
hermeticity, we obtain
\begin{equation}
\int_0^\infty\int_{-\infty}^{\infty}a^{2-3\alpha}\Psi^*\Psi\,da\,d\beta_+ 
= \frac{2\sqrt{2\gamma\pi}}{3\lambda(1 - \alpha)}F(T)
\end{equation}
where
\begin{equation}
F(T) = \exp{(\frac{C_I^2}{2\gamma})} \quad ,
\end{equation}
and
\begin{eqnarray}
&C&(a,B) = C_R + iC_I \quad ,\nonumber\\ 
&C&_R = \ln a - \frac{1}{3(1 - \alpha)}\ln 4B^*B \quad ,\nonumber\\
&C&_I = \frac{- 2}{3(1 - \alpha)}\tan^{-1}(\frac{sT}{\lambda})
\quad .
\end{eqnarray}
The norm of the wave function is time dependent. Hence, the quantum
model is not unitary. 
\par
The absence of unitarity may be understood by inspecting again the
wave packet (\ref{wp}). In fact, this wave packet goes to zero at infinity,
excepted along the line $\beta_0 = - \beta_+$, where it takes a constant value
at infinity. This does not spoil the regularity of the wave packet; in particular,
it remains finite when integrated in all space and
specific boundary conditions are obeyed. But, this leads, at the
same time, to an anomaly at the infinity boundary. The reason for this anomaly may be understood by analysing again
the Schr\"odinger-like equation (\ref{wdwe}). Notice that, after decomposing it into
stationary states, the energy $E$ is zero along the whole line $\beta_0 = - \beta_+$.
Along this line, the wave function need not to vanish.
\par
It would be expected that a hermitian Hamiltonian operator should always lead to
a unitary quantum system, since the Hamiltonian operator is responsible for the
time evolution of the quantum states. The problem here relies on the fact that, in
spite of being hermitian, the Hamiltonian effective operator
\begin{equation}
\label{eh}
H_{eff} = e^{-3(1 - \alpha)}\biggr\{\partial_0^2 - \partial_+^2 - \partial_-^2\biggl\}
\end{equation}
is not self-adjoint. This
means that $H^\dag = H$ but the domain of $H^\dag$ is not the same as the domain of
$H$, and the conservation of the norm becomes senseless\cite{symon}.
\par
In order to verify if an operator is self-adjoint or not, we must compute the so-called
deficiency indices $n_\pm$ which are the dimensions of the linear independent square integrable
solutions of the indicial equation
\begin{equation}
H\phi = \pm i\phi \quad .
\end{equation}
Using the effective Hamiltonian (\ref{eh}), the solutions of the indicial equations are
\begin{eqnarray}
\phi_+ = c_1J_\nu(y) + c_2J_{-\nu}(y) \quad , \\
\phi_- = c_3K_\nu(y) + c_4I_\nu(y) \quad ,
\end{eqnarray}
where $\nu = ik/r$ and $y = \sqrt{i/r^2}\,a^r$, $a$ and $r$ having the same definitions as
before. It is easy to see that $J_{\pm\nu}(y)$ and $I_\nu(y)$ are not square integrable solutions while $K_\nu(y)$ is. Hence, $n_+ = 0$ and $n_- = 1$ and, as explained in\cite{symon}, the effective Hamiltonian operator is not self-adjoint and does not
admit any self-adjoint extension. Notice that changing arbitrarily the signature
in (\ref{eh}) or suppressing the unusual measure, the deficiency indices become
$n_+ = n_- = 0$ and the effective Hamiltonian becomes self-adjoint.
\par
Since the anisotropic quantum cosmological model with perfect fluid studied here is
non-unitary, we expect that the usual many-worlds\cite{tipler} and ontological (dBB)\cite{holland} interpretations
of quantum mechanics give different results. In fact, computing the expectation values
for the scale factor functions $\beta_i$ we obtain,
\begin{equation}
<\beta_0> = \frac{1}{3(1 - \alpha)}\biggr\{\ln\biggr(\frac{2\vert B\vert^2}{\lambda}\biggl) + n\biggl\}
\quad .
\end{equation}
\begin{equation}
<\beta_+> = \frac{1}{3(1 - \alpha)}\biggr\{\ln(2\lambda) - n\biggl\} = \mbox{const.} \quad .
\end{equation}
These results lead to an isotropic universe.
\par
The computation of the bohmian trajectories, on the other hand, leads to the following
results:
\begin{eqnarray}
e^{\beta_0} &=& \biggr(\frac{-1}{24s\lambda\gamma}\biggl)^\frac{1}{3(1 - \alpha)}\biggr[\lambda^2 + s^2T^2\biggl]^\frac{1}{3(1 - \alpha)}\times\nonumber\\
&\times&\biggr[(\tan^{-1})^2\biggr(\frac{sT}{\lambda}\biggl) 
+ E\biggl]^\frac{1}{3(1 - \alpha)} \quad , \\
\beta_+ &=& - \frac{1}{3(1 - \alpha)}\ln\biggr\{(\tan^{-1})^2\biggr(\frac{sT}{\lambda}\biggl) + E\biggl\} \nonumber\\
&+& \ln\biggr\{\biggr[-24s\lambda\gamma\biggl]^\frac{1}{3(1- \alpha)}\biggl\} + \ln D
\quad ,
\end{eqnarray}
where $E$ and $D$ are integration constants. Remember that $s < 0$.
\par
In opposition to the expressions obtained for the expectation values of $\beta_0$ (which is connected to $a$)
and $\beta_+$ in the preceding subsection, the bohmian trajectories predict an anisotropic universe.

\vspace{0.2cm}
{\bf Acknowledgements:} We thank N. Pinto-Neto, J. Acacio de Barros and N.A. Lemos for many enlightfull
discussions and CNPq (Brazil) for partial financial support.

\end{document}